\documentclass[%
 floatfix,
 reprint,
superscriptaddress,
 amsmath,amssymb,
 aps,
 twocolumn,
float,prl]{revtex4-2}

\usepackage{graphicx}
\usepackage{dcolumn}
\usepackage{bm}
\usepackage{xcolor}
\usepackage[caption=false]{subfig}
\usepackage{tabularx}
\usepackage{epstopdf}

\begin{document}

\newcommand{\Schro}{Schr\"{o}dinger~}
\newcommand{\ket}[2]{\left|#1,#2\right\rangle} 
\newcommand{\bra}[2]{\left\langle#1,#2\right|} 
\newcommand{\me}[4]{\left\langle#1,#2\right|\tilde{\rho}\left|#3,#4\right\rangle} 

\title{One-Positron Quantum Cyclotron}

\author{T. G. Myers}
  \affiliation{Center for Fundamental Physics, Department of Physics and Astronomy, Northwestern University, Evanston, Illinois 60208, USA}

\author{L. Soucy}
  \affiliation{Center for Fundamental Physics, Department of Physics and Astronomy, Northwestern University, Evanston, Illinois 60208, USA}
  
 \author{B. A. D. Sukra}
  \affiliation{Center for Fundamental Physics, Department of Physics and Astronomy, Northwestern University, Evanston, Illinois 60208, USA}

   \author{B. Sinha}
  \affiliation{Center for Fundamental Physics, Department of Physics and Astronomy, Northwestern University, Evanston, Illinois 60208, USA}
 
\author{G. Gabrielse}
 \email{gerald.gabrielse@northwestern.edu}
 \affiliation{Center for Fundamental Physics, Department of Physics and Astronomy, Northwestern University, Evanston, Illinois 60208, USA}
\date{\today}

\newcounter{gg}
\newenvironment{CompactEnumerate}{
\begin{list}{\arabic{gg}.}{\usecounter{gg} \setlength{\topsep}{-2pt} \setlength{\rightmargin}{30pt} \setlength{\itemsep}{-5pt}}}{\end{list}\smallskip}

\newcommand{\nuap}{\nu_a^\prime}
\newcommand{\nucp}{\nu_c^\prime}
\newcommand{\nuab}{\bar{\nu}_a}
\newcommand{\nucb}{\bar{\nu}_c}
\newcommand{\nuzb}{\bar{\nu}_z}
\newcommand{\numb}{\bar{\nu}_m}
\newcommand{\fcb}{\bar{f}_c}
\newcommand{\zhat}{\bf \hat{z}}

\begin{abstract}     
A one-positron quantum cyclotron is realized with a single positron suspended indefinitely in the magnetic field of a Penning trap.  This opens the possibility of quantum measurements of the positron magnetic moment at a precision much higher than attained with classical cyclotron motion.  Comparing the magnetic moments measured using positron and electron quantum cyclotrons should provide the most stringent test of the fundamental CPT invariance of the Standard Model of particle physics in the lepton sector.  
\end{abstract}

\pacs{13.40.Em, 14.60.Cd, 12.20-m}

\maketitle

\begin{figure}[htbp!]
\centering
\includegraphics[width=0.95\columnwidth]{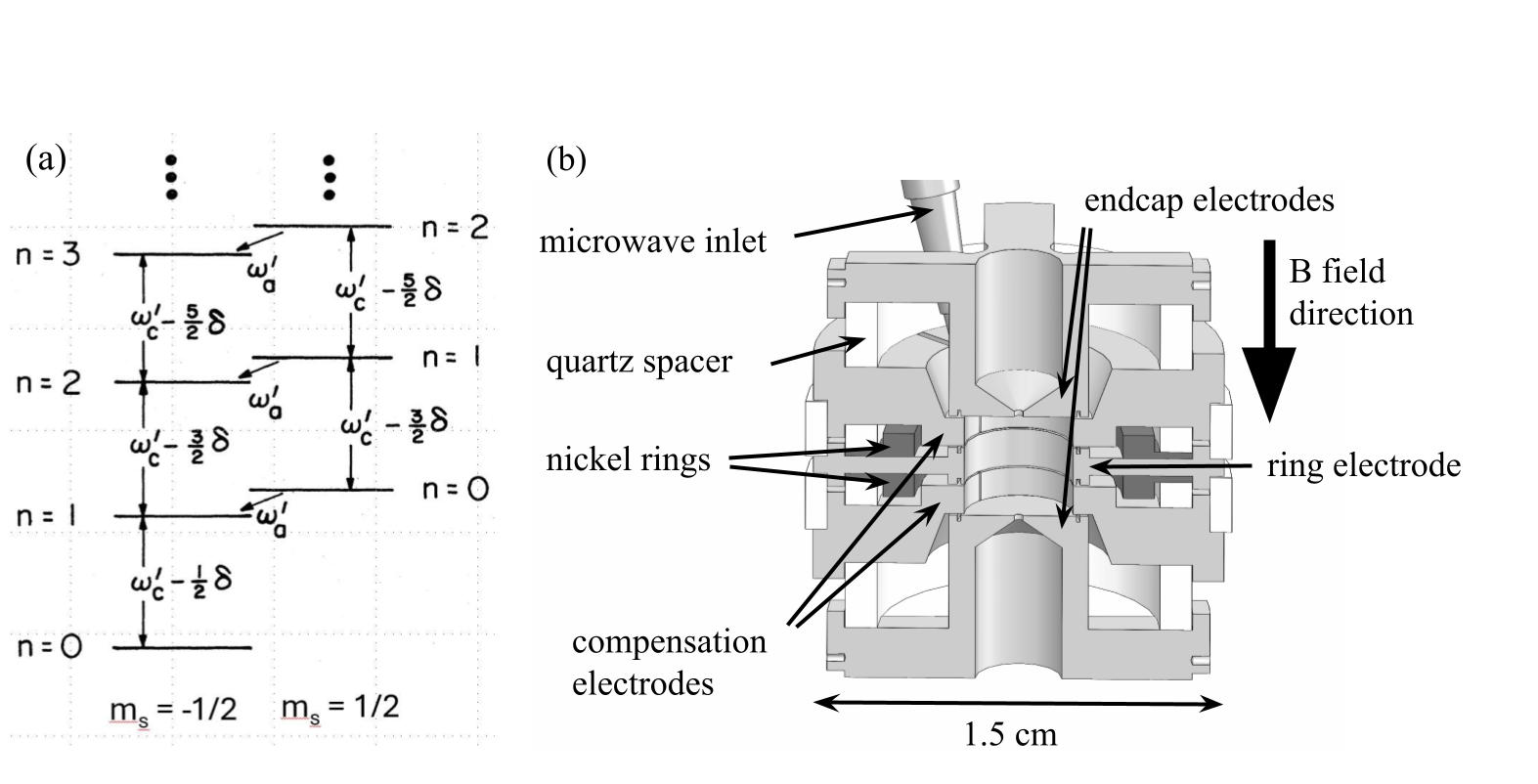}
\caption{(a) Energy levels of a one-positron quantum cyclotron. (b) Trap electrodes that confine the positron.}
\label{fig:LevelsAndTrap}
\end{figure}

A one-positron quantum cyclotron is realized for the first time.  This is a single positron (charge e and mass m) suspended in a magnetic field with its lowest cyclotron (and spin) energy levels fully resolved. The levels (Fig.~\ref{fig:LevelsAndTrap}a) are separated in this demonstration using a $B=5.6$ T magnetic field from a 4.2 K superconducting solenoid. The cyclotron ground state is initially prepared by bringing this motion into thermal equilibrium with the surrounding trap electrodes (Fig.~\ref{fig:LevelsAndTrap}b), within a completely sealed vacuum enclosure, all kept at 300 mK using a dilution refrigerator. The cold trap and enclosure essentially eliminate blackbody photons that would excite cyclotron states \cite{QuantumCyclotron}.  They also serve as a cryopump that produces a nearly perfect vacuum \cite{PbarMass}, much better than is needed to store a positron as long as desired without it annihilating in a collision with a background gas atom. Quantum nondemolition (QND) detection of the positron cyclotron and spin states is also critical, to measure these quantum states without changing them.  A significant challenge is the need to slow, capture and cool a beta emission from a radioactive $^{22}$Na nucleus (with a kinetic energy of up to 546 keV) until its thermal cyclotron energy is only 26 $\mu$eV. The positron path into a high-precision Penning trap must allow a useful positron loading rate, without compromising the trap cavity's ability to inhibit spontaneous emission
\cite{InhibitionLetter}, since this provides the averaging time essential for determining the cyclotron state.

To date, the most precise positron magnetic moment measurement was made using classical cyclotron motion \cite{DehmeltMagneticMoment}. 
The one-positron quantum cyclotron should open the way to quantum magnetic moment measurements with much higher precisions. A one-electron quantum cyclotron \cite{QuantumCyclotron} made is possible to measure the electron magnetic moment to $1$ part in $10^{13}$ \cite{NorthwesternMagneticMoment2023}, for example. This most precisely determined property of any elementary particle was measured in order to test the Standard Model’s most precise prediction \cite{aliberti2025anomalousmagneticmomentmuon}. A quantum positron magnetic moment measurement at this precision would provide the most direct and precise test of the fundamental CPT invariance of the Standard Model in the lepton sector. For comparison, classical cyclotron motion of protons and antiprotons is used to test baryon CPT invariance \cite{PbarMagneticMoment} much less precisely, most recently to $1$ part in $10^9$ \cite{PbarMagneticMomentBase2017}.  Meson CPT invariance is tested much more precisely by measuring asymmetries in the decay of long and short lived kaons \cite{kaon}.

The cyclotron and spin energy levels of a one-positron quantum cyclotron (Fig.~\ref{fig:LevelsAndTrap}a) are given by
\begin{equation}
E_{n m_s} = \hbar \omega_c (n + \tfrac{1}{2}) +\hbar \omega_s m_s, 
\end{equation}
with a cyclotron quantum number ($n=0,1,...$) and a spin quantum number ($m_s=\pm1/2$).  Choosing $\zhat$ to be the direction opposite to the magnetic field makes $\omega_c=eB/m$ be the positive angular cyclotron frequency for a classical cyclotron orbit of a positron. The cyclotron levels are then equally-spaced by $\hbar \, \omega_c$, where $\hbar$ is Planck's constant. (The figure also represents a tiny relativistic shift $\delta$ that can be neglected for this demonstration, given that $\delta/\omega_c \approx 10^{-9}$.)  Fig.~\ref{fig:LevelsAndTrap}a shows two cyclotron ``ladders'', the left for spin down ($m_s=-1/2$) and the right for spin up ($m_s=1/2$).  The spin levels are spaced by the spin precession frequency $\omega_s = (g/2)\, \omega_c$.  The frequency difference $\omega_a=\omega_s - \omega_c$, called the anomaly frequency for historical reasons, is also represented in the figure.  All three frequencies are proportional to a magnetic field from a superconducting solenoid that drifts by only 2 ppb/day  \cite{Helium3NMR2019}.  

The proportionality constant between the spin and cyclotron frequencies, $g/2 = \mu/\mu_B$, is the spin magnetic moment $\mu$ in units of Bohr magnetons, $\mu_B = e\hbar /2 m $.  This dimensionless magnetic moment is larger than unity by only about 1 part in a thousand. With positron energy levels  unambiguously resolved, quantum jump spectroscopy could be used to measure two of the frequencies, and thus determine the dimensionless magnetic moment for the positron.

For $B=5.6$ T, the cyclotron and spin frequencies are both approximately $156$ GHz, differing only by a part in a thousand.  A positron is trapped by adding an electrostatic quadrupole potential $V \propto z^2 - \rho^2/2$  (with $\boldsymbol{\rho} \equiv 
x\mathbf{\hat{x}}+y\mathbf{\hat{y}}$) to $B\hat{z}$  \cite{Review}.  The electrodes of a cylindrical Penning trap \cite{CylindricalPenningTrap,CylindricalPenningTrapDemonstrated} (Fig.~\ref{fig:LevelsAndTrap}b), with appropriately chosen relative dimensions and applied potentials, produces such a potential.  It causes a  positron on the central axis to oscillate along $\hat{z}$ at the axial frequency $\omega_z/2\pi \approx 206$ MHz. We presently discuss how small changes in this frequency are used to detect changes in the cyclotron and spin states.  A circular magnetron motion perpendicular to the magnetic field \cite{Review}, at about $135$ kHz, also arises from the addition of the electrostatic trapping potential.  The magnetron motion is cooled by axial sideband cooling  \cite{Review} so that its effect is not important for this realization. The addition of the electrostatic quadrupole potential also slightly shifts the cyclotron frequency in a well-understood way \cite{Review}, from $\omega_c$ to $\omega_c^\prime$ in Fig.~\ref{fig:LevelsAndTrap}a. This small shift has no significant consequence for this realization of the one-positron quantum cyclotron so we take $\omega_c^\prime \approx \omega_c$ in what follows.

\begin{figure}[htbp!]
\centering
\includegraphics[width=0.6\columnwidth]{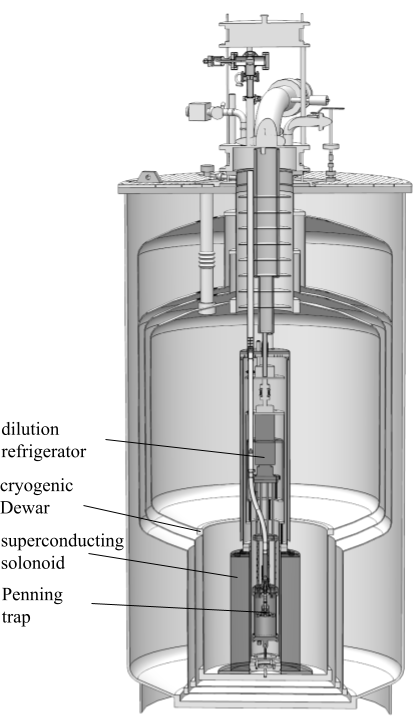}
  \caption{The cryogenic dewars, superconducting solenoid, dilution refrigerator, and Penning trap.}
\label{fig:WholeApparatus}
\end{figure}

\begin{figure}[htbp!]
\centering
\includegraphics[width=0.42\columnwidth]{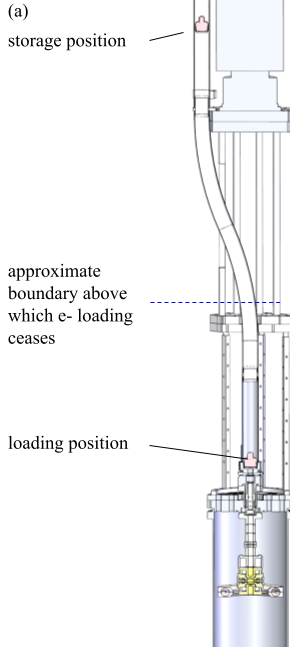}
\includegraphics[width=0.505\columnwidth]{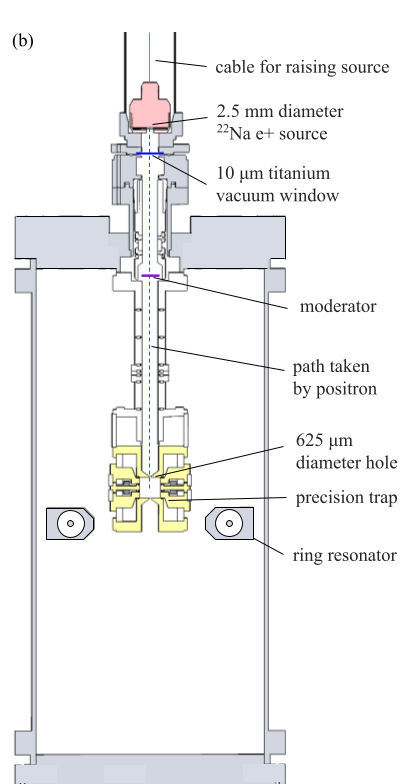}
   \caption{(a) Positron source locations.  (b) Positron path from a radioactive source into a highly-harmonic cylindrical Penning trap with flat endcap electrodes.}  
   \label{fig:PositronLoading}
\end{figure}

The cryogenic apparatus used to realize the one-positron quantum cyclotron is a cold-bore dewar system (Fig.~\ref{fig:WholeApparatus})  containing a persistent 4.2 K superconducting solenoid that produces the 5.6 T magnetic field. A dilution refrigerator apparatus is lowered into it from above.  The fridge cools the trap electrodes and the surrounding vacuum enclosure that hangs beneath it \cite{EfficientPositronAccumulation} to about 300 mK.  However, both the trap and its enclosure rest mechanically upon the 4.2 K aluminum tube upon which the superconducting solenoid is wound. This apparatus differs from what was used to measure the electron magnetic moment \cite{NorthwesternMagneticMoment2023} most notably in that a radioactive $^{22}$Na positron source can be lowered on a thin nylon cable from room temperature, down through apertures in the dilution refrigerator (Fig.~\ref{fig:PositronLoading}a) until it settles at the top of the vacuum enclosure (Fig.~\ref{fig:PositronLoading}b). Once a positron (or electron) is loaded into the trap, the source is raised to the indicated storage position to prevent the loading of additional particles.

The positron loading setup in Fig.~\ref{fig:PositronLoading} is the simplest method devised for accumulating positrons directly in an ultrahigh, cryogenic vacuum. Its straightforward construction, wiring, and operation make it well-suited to precision measurements. The tradeoff that slows initial trap optimization is the 8-11 hours required to load a positron.  This is very slow per mCi compared to more elaborate implementations  \cite{PositronsFromPositronium,EfficientPositronAccumulation}.  If needed,  more rapid positron accumulation, enhanced control, and better diagnostics could be attained if the grounded cylindrical electrodes mounted just above the precision trap would be configured as a positron accumulation trap. 
Accumulated positrons could then be transferred from into the precision trap.  Slow accumulation suffices for this demonstration because once the potentials applied to the high precision trap are tuned well enough that the axial oscillation of a single positron is easily detectable, positron loading is not often required.  A single positron has been trapped for months. 

The positrons originate as the beta emissions of $^{22}$Na nuclei within a completely sealed source. Despite the high energy spectrum endpoint of 546 keV, an efficient method to accumulate positrons  nearly at rest \cite{PositronsFromPositronium, EfficientPositronAccumulation}, summarized below,  makes it possible to use a very weak 18 $\mu$Ci source. This is of practical importance because modern safety practices in a university laboratory would make it extremely difficult to implement a much larger source, such as the 500 $\mu$Ci source installed for the best classical cyclotron measurement of the positron magnetic moment \cite{UwPositrons}.

Fig.~\ref{fig:PositronLoading} represents the path of positrons (green dashes) from a source (red), through a 10 $\mu m$ thick titanium window (blue) into a completely enclosed vacuum container (gray) that is kept at about 300 mK by a thermal connection to a dilution refrigerator. The positrons lose energy via collisions within a single crystal tungsten (100) moderator (purple) \cite{ModerationReview}. In accordance with earlier studies \cite{PositronsFromPositronium,EfficientPositronAccumulation}, slowed positrons near the exit side of the moderator experience the work function of the tungsten, which has a sign that ejects very slow positrons out of the moderator. A positron  that proceeds downward along a vertical magnetic field line can attract an electron from the tungsten surface, and both particles are guided along parallel magnetic field lines.  The strong field keeps the positron and electron loosely bound in a highly-magnetized Rydberg positronium atom that  travels through a very small (625 $\mu$m) hole in a silver trap electrode into the interior of a cylindrical Penning trap (represented in gold) built for positron studies.  
The weakly-bound Rydberg system proceeds unaffected by the trapping potential of the biased trap electrodes because it is neutral, until the electric field becomes large enough to ionize the highly-magnetized Rydberg positronium within the trap.  The negative electron is lost, but the positive positron is captured in the potential well of the trap.  The hole diameter must be sufficiently large to allow positron loading without ionizing the positronium before it is within the trapping well.  The diameter must be sufficiently small that the quality factor of the low-loss microwave cavity remains high enough to substantially suppress the spontaneous emission of synchrotron radiation.  The inhibition is essential to provide the averaging time needed to resolve excited cyclotron states before they decay.

Only a positive particle can be trapped in the electrostatic quadrupole that is applied. The presence of a positron in the trap is confirmed by the $e/m$ ratio deduced by measuring its axial frequency
\begin{equation} 
\omega_z = \sqrt{ a ({e}/{m})}.
\end{equation}
The trap constant $a$ is well calculated and understood \cite{CylindricalPenningTrap}. It is linear in the trapping voltage applied to the electrodes and depends upon the well-known size and shape of these electrodes.

\begin{figure}[htbp!]
\centering
\includegraphics[width = \columnwidth]{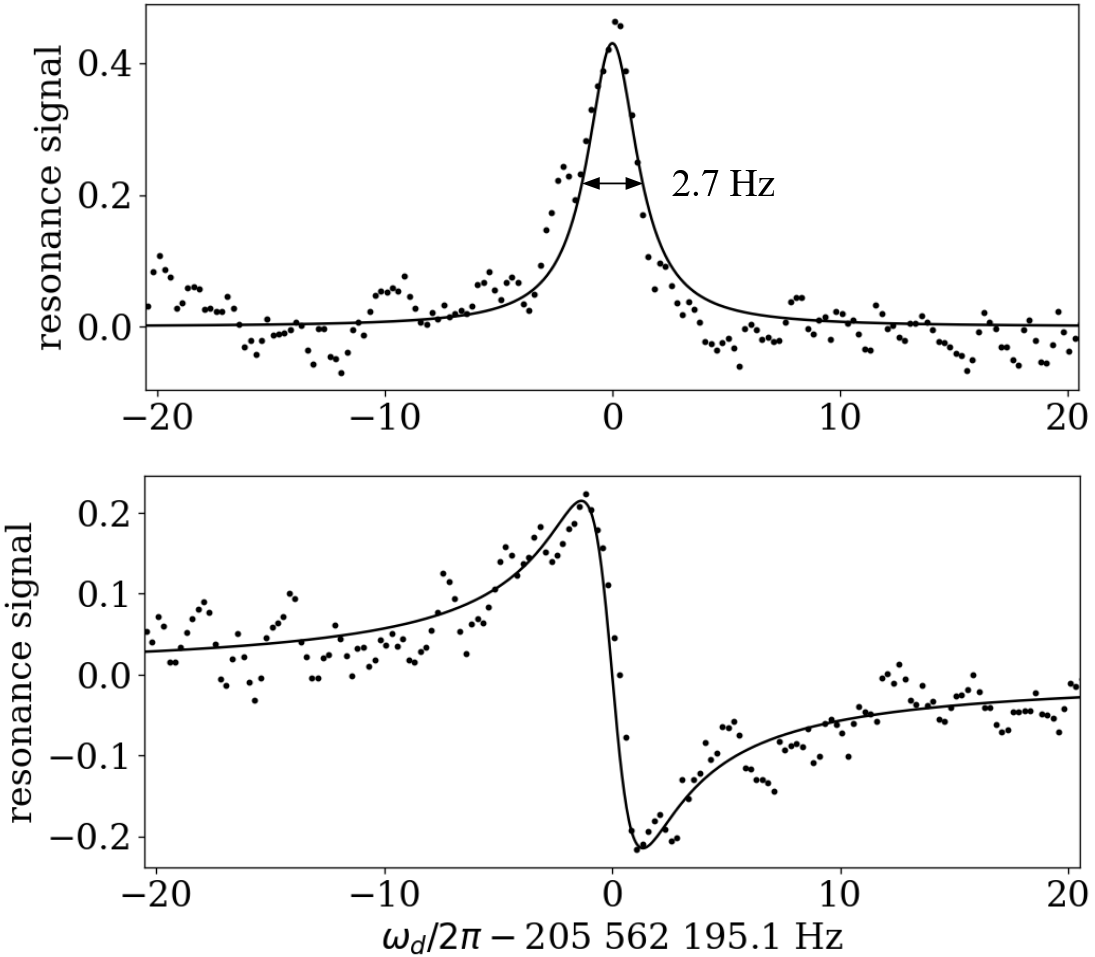}
 \caption{Driven axial resonances of a single positron (solid) with best fit Lorentzian lineshapes.}
   \label{fig:AxialResonance}
\end{figure}

To measure $\omega_z$, the axial damping rate $\gamma_z$, and the number of trapped positrons, the axial oscillation of trapped positrons along the magnetic field direction is driven with an oscillating potential applied to the trap electrodes.  The  frequency of the drive $\omega_d$ is swept through resonance with $\omega_z$. Fig.~\ref{fig:AxialResonance} shows the observed signal from the oscillating positron charge that is in phase (Fig.~\ref{fig:AxialResonance}a) and out of phase (Fig.~\ref{fig:AxialResonance}b) with the drive.  The drive strength is very weak in this illustration to minimize power shifts and broadening, so the signal-to-noise ratio is much smaller than for a stronger drive.  The oscillating charge induces a current through a resistance $R$ that is part of a resonant LCR circuit (with a Q of 900) attached to the trap electrodes to keep the trap capacitance from shorting out the induced signal.  

A coaxial resonator, formed into a circle around the electrodes (Fig.~\ref{fig:PositronLoading}) is the central component of the LCR circuit.  This resonator geometry, not previously used for such particle detection, has its inner and outer conductors  shorted together at one end of the resonator and connected to a pair of adjacent trap electrodes at the other. The circuit is the input matching network of a cryogenic HEMT (high-electron-mobility transistor) amplifier.  Its output is amplified further with higher temperature amplifiers that follow it, and the amplified signal is digitized. 

 Power dissipated in R damps the positron's axial motion and determines the widths of the observed axial resonance signals. Simultaneous fits to the expected line shapes in Fig.~\ref{fig:AxialResonance}a and Fig.~\ref{fig:AxialResonance}b  reveal the axial frequency and the damping rate $\gamma_z/2\pi \approx 2.7$ Hz, which corresponds to $R = 80\,\rm{k}\Omega$.  For the center-of-mass axial motion of more than one positron, the damping (and hence the observed signal width) increases in proportion to the number of particles. A confirmation that only one positron is trapped comes from spin flip observations that are discussed presently.

Small shifts in $\omega_z$ unambiguously resolve one-quantum spin and cyclotron jumps.  Saturated nickel rings encircling the trap (Fig.~\ref{fig:LevelsAndTrap}b) produce a  magnetic bottle gradient, $\Delta \vec{B}=B_2\left[(z^2 - \rho^2/2)\hat{z}-z\rho\hat{\rho}\right]$, and $B_2 = 1500~\rm{T}/\rm{m}^2$ is calculated \cite{Review}.  This  
gradient shifts $\omega_z$ in proportion to the cyclotron and spin quantum numbers, 
\begin{equation}
\Delta \omega_z = \beta \, (n + m_s),
\end{equation}
with $\beta/2\pi \approx 3.8$ Hz for one positron.  
(If the axial center-of-mass motion of two trapped positrons is detected, $\beta/2\pi$ is half this value since $\beta$  scales inversely with the total mass of the observed particles \cite{Review}.)   
A measured shift is a quantum nondemolition (QND) measurement of the cyclotron and spin states that reveals the state of the quantum cyclotron without this detection itself changing the spin and cyclotron states.  

The shift $\Delta \omega_z$ must be measured before the cyclotron state decays via the spontaneous emission of synchrotron radiation.  Without trap electrodes, the $n=1$ cyclotron state decays via electric dipole radiation at a rate \cite{Review}
\begin{equation}
    \gamma_c = \frac{4}{3} \alpha \frac{\hbar \omega_c}{mc^2} \omega_c,
\end{equation}
where $\alpha$ is the fine structure constant.
At 5.6 T, this corresponds to a  lifetime of only $\gamma_c^{-1}=83$ ms. The small signal from only one trapped positron must be averaged for longer than this decay time to allow measuring $\Delta \omega_z$.  It is thus critical for realizing a one-positron quantum cyclotron that we inhibit this spontaneous emission rate using trap electrodes that also serve as a low-loss microwave cavity with the positron suspended at its center.  The spontaneous emission rate is observably reduced by up to a factor of 200 or more, depending upon how far away $\omega_c$ is from the frequency of coupled cavity radiation modes \cite{InhibitionLetter}.  For this realization, $\gamma_c^{-1} \approx 4$ s.  

Blackbody photons that could excite $n=0$ to $n=1$ are essentially not present when the trap cavity is at 300 mK.  
The Boltzmann probability that the $n=1$ cyclotron state is thermally populated is only  slightly larger than a part in $10^{11}$, given that the cyclotron energy levels are spaced by about $\hbar \omega_c/k_B \approx 7.5$ K in temperature units   \cite{QuantumCyclotron}.

\begin{figure}[htbp!]
\centering
\includegraphics[width = \columnwidth]{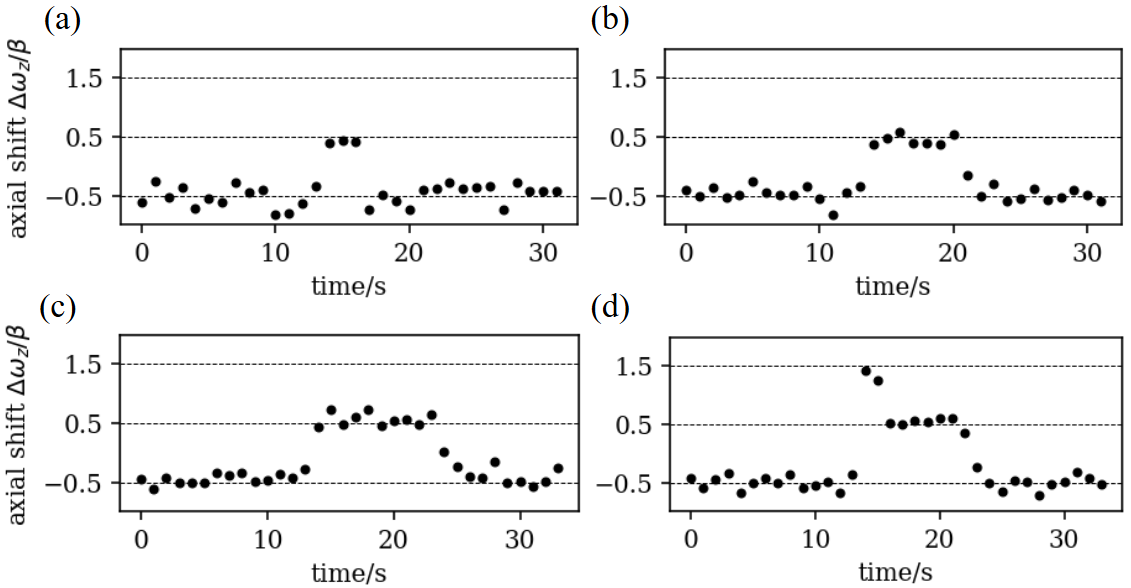}
 \caption{QND measurements of the cyclotron energy of a one-positron quantum cyclotron illustrate cyclotron excitations driven by a nearly resonant microwave drive, followed by energy decays via the spontaneous emission of synchrotron radiation.}
   \label{fig:CyclotronJumps}
\end{figure}

 A self-excited single electron oscillator \cite{SelfExcitedOscillator} was used for measurements of the electron magnetic moment \cite{NorthwesternMagneticMoment2023} to observe the small $\Delta \omega_z$ shift 
 before a cyclotron excited state decays. For this first demonstration of positron quantum jumps, the simpler scheme employed is sweeping the frequency of an axial driving force (much stronger than used for Fig.~\ref{fig:AxialResonance})  through resonance to impulsively excite an axial oscillation while continuously Fourier-transforming the induced signal. 
 Fig.~\ref{fig:CyclotronJumps} illustrates how $\omega_z$, and hence the cyclotron energy, change as the positron is first excited from its ground state and then subsequently decays.  The cyclotron motion is excited by a 156 GHz microwave drive introduced into the trap at a frequency near to the cyclotron frequency $\omega_c$.  The one-quantum cyclotron excitations from a cyclotron ground state in Fig.~\ref{fig:CyclotronJumps}a-c subsequently decay via the inhibited spontaneous emission of synchrotron radiation. For a stronger cyclotron drive,  Fig.~\ref{fig:CyclotronJumps}d shows an example where the cyclotron energy increases by two quanta in quick succession, followed by two spontaneous emissions. The clear energy steps illustrate the realization of a one-positron quantum cyclotron.

\begin{figure}[htbp!]
\centering
\includegraphics[width = 0.70\columnwidth]{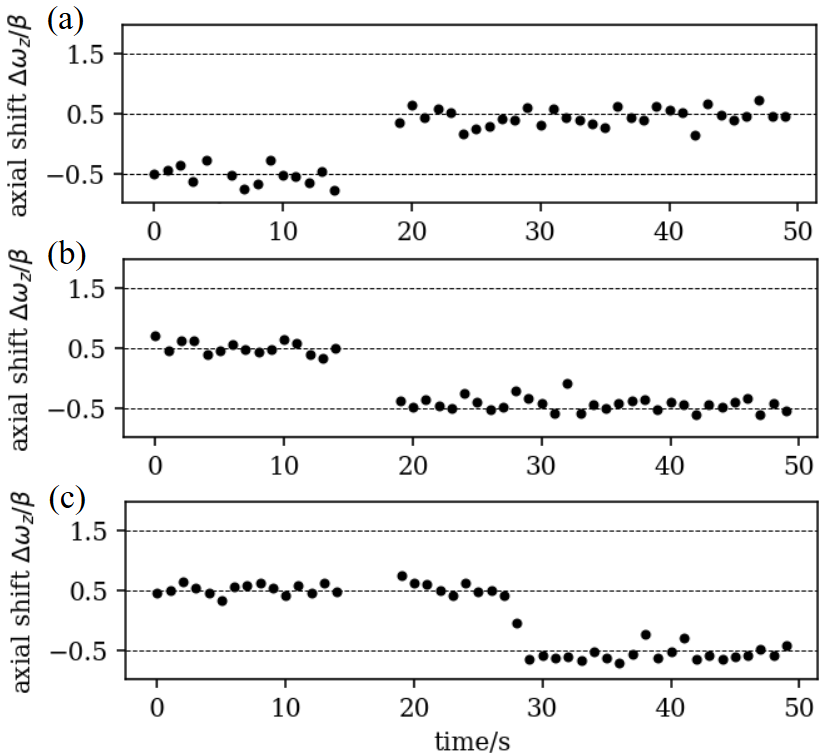}
 \caption{QND observations of spin flips caused by simultaneously applied anomaly and cyclotron drives in (a), and by a driven anomaly transition followed by a spontaneous emission in (b) and (c).  }
\label{fig:SpinFlips}
\end{figure}

The spin state of the one-positron quantum cyclotron is much less coupled to its environment \cite{Review}. The cyclotron ground state ($n=0$) with spin down ($m_s=-1/2$) does not decay.  The cyclotron ground state ($n=0$) with $m_s=1/2$ decays via a magnetic dipole transition to the spin down ground state at a rate \cite{Review}
\begin{equation}
\gamma_s \approx \frac{\hbar \omega_s}{mc^2} \gamma_c
\end{equation}
For B = 5.6 T, the spin decay rate is about 9 orders of magnitude smaller than the cyclotron decay rate, corresponding to a calculated decay time $\gamma_s^{-1} = 2.1$ years. The trap microwave cavity likely suppresses this rate further.
The lowest cyclotron states for both spin directions are thus  effectively stable ground states and no inhibited emission is needed to resolve the spin state.  

Fig.~\ref{fig:SpinFlips} gives examples of QND observations of driven spin flips.  In all three examples, an oscillatory potential at a frequency close to $\omega_a$ is applied to the trap electrodes.  This drives a far off-resonance axial oscillation of the positron through the radial magnetic gradient $B_2z \boldsymbol{\rho}$ from the magnetic bottle rings.  The positron sees the oscillating magnetic field perpendicular to $\hat{z}$ as needed to flip its spin, with a radial gradient that allows a simultaneous cyclotron transition \cite{Review}. The axial frequency shift is not measured during 2 seconds that the anomaly drive is applied, nor during the seconds before and after. 

Fig.~\ref{fig:SpinFlips}a shows the spin being flipped from down to up, from ($n=0$, $m_s=-1/2$) to ($n=0$, $m_s=1/2$) via a two step process. A cyclotron drive that is near to resonance with $\omega_c$ first excites the intermediate state ($n=1$, $m_s=-1/2$).  The simultaneously applied anomaly drive near to resonance at $\omega_a$ then causes a transition to ($n=0$, $m_s=1/2$).  
Fig.~\ref{fig:SpinFlips}b-c shows the spin being flipped from up to down, from ($n=0$, $m_s=1/2$) to ($n=0$, $m_s=-1/2$).  These are also two-step processes.  The 
near-resonance anomaly drive changes the state from ($n=0$, $m_s=1/2$) to ($n=1$, $m_s=-1/2$).  This state subsequently decays to ($n=0$, $m_s=-1/2$) via spontaneous emission. The cyclotron excited state decays quickly in Fig.~\ref{fig:SpinFlips}b, and well after the anomaly transition in Fig.~\ref{fig:SpinFlips}c.

The observed size of the axial frequency change for all cyclotron jumps and spin flips, $\beta/2\pi \approx 3.8$ Hz, confirms that only one positron is being observed.  If the center-of-mass motion of 2 positrons was being observed, $\beta$ would be half this value. The observed shift of 3.8 Hz could be observed if both spins flip, but on some trials only one spin would flip and the observed shift would be half what is observed.  No shift would be observed if one of the spins flips from down to up while the other flips from up to down.

In summary, a one-positron quantum cyclotron is realized for the first time.    A positron from the decay of a $^{22}$Na nucleus, with an endpoint energy of 546 keV, is trapped in a Penning trap cooled to about 300 mK by a dilution refrigerator.  The slowed, cooled and confined positron occupies only its cyclotron ground state until it is excited by a resonant driving force. A trap cavity is demonstrated that provides positron access into the trap without compromising the cavity-inhibited spontaneous emission that is critical for observing cyclotron excitations before they decay.   Quantum nondemolition (QND)  detection reveals the quantum cyclotron and spin states without changing either. 

With further optimization, and the considerable time it takes to investigate uncertainties, it now seems feasible to make a quantum measurement of the positron magnetic moment 
at the precision of the most recent electron magnetic moment measurement.  This would test the fundamental CPT invariance of the Standard Model more than 30 times more precisely than has been possible in the lepton sector.

\begin{acknowledgments}
We are grateful to G.\ Nahal for his help in the design and construction of the coaxial resonator.  This work was supported by the National Science Foundation award number 2110565.   
\end{acknowledgments}
 

\end{document}